%% file: main.tex
\documentclass[sigconf, screen]{acmart}

\setcopyright{none} 
\renewcommand\footnotetextcopyrightpermission[1]{} 

\input{packages}
\input{macros}

\copyrightyear{2026}
\acmYear{2026}
\setcopyright{cc}
\setcctype{by}
\acmConference[AST '26]{7th ACM/IEEE International Conference on Automation of Software Test (AST 2026)}{April 13--14, 2026}{Rio de Janeiro, Brazil}
\acmBooktitle{7th ACM/IEEE International Conference on Automation of Software Test (AST 2026) (AST '26), April 13--14, 2026, Rio de Janeiro, Brazil}
\acmPrice{}
\acmDOI{10.1145/3793654.3793744}
\acmISBN{979-8-4007-2476-3/2026/04}

\begin{document}



\title{Improving Deep Learning Library Testing with Machine Learning}

\author{Facundo Molina}
\authornote{Work done while at the IMDEA Software Institute.}
\orcid{0000-0002-2441-1555}
\affiliation{
  \institution{Complutense University of Madrid}
  \city{Madrid}
  \country{Spain}
}
\email{facundom@ucm.es}

\author{M M Abid Naziri}
\orcid{0000-0002-3499-5283}
\affiliation{
  \institution{North Carolina State University}
  \city{Raleigh}
  \country{USA}
}
\email{mnaziri@ncsu.edu}

\author{Feiran Qin}
\orcid{0009-0003-9310-0798}
\affiliation{
  \institution{North Carolina State University}
  \city{Raleigh}
  \country{USA}
}
\email{fqin2@ncsu.edu}

\author{Alessandra Gorla}
\orcid{0000-0002-6711-3080}
\affiliation{
  \institution{IMDEA Software Institute}
  \city{Madrid}
  \country{Spain}
}
\email{alessandra.gorla@imdea.org}

\author{Marcelo d'Amorim}
\orcid{0000-0002-1323-8769}
\affiliation{
  \institution{North Carolina State University}
  \city{Raleigh}
  \country{USA}
}
\email{mdamori@ncsu.edu}

\begin{abstract}
Deep Learning (DL) libraries like \tf{} and \torch{} simplify machine
learning (ML) model development but are prone to bugs due to their
complex design. Bug-finding techniques exist, but without precise API
specifications, they produce many false alarms. Existing methods to
mine API specifications lack accuracy.

We explore using ML classifiers to determine input validity. We
hypothesize that tensor shapes are a precise abstraction to encode
concrete inputs and capture relationships of the data. Shape
abstraction severely reduces problem dimensionality, which is
important to facilitate ML training. Labeled data are obtained by
observing runtime outcomes on a sample of inputs and classifiers are
trained on sets of labeled inputs to capture API constraints.
  
Our evaluation, conducted over 183 APIs 
from \tf{} and \torch{}, shows that 
the classifiers generalize well on unseen data with over 91\%
accuracy. Integrating these classifiers into the pipeline of ACETest,
a \sota{} bug-finding technique, improves its pass rate from
$\sim$29\% to $\sim$61\%.
Our findings suggest that
ML-enhanced input classification is an important aid to scale DL
library testing.

\end{abstract}

\maketitle

\section{Introduction}






Deep Learning (DL) has become a cornerstone of modern computation,
revolutionizing fields such as image and text generation.
In Software Engineering, DL has been used to automate tasks such as
code generation, refactoring, and analysis. Due to the complexity of
developing DL pipelines from scratch, many software projects rely on
well-established deep learning libraries, such as
\tf{}~\cite{abadi2016tensorflow} and
\torch{}~\cite{paszke2019pytorch}. These libraries abstract low-level
implementation details, allowing developers to focus on higher-level
design and application logic. However, these libraries are complex and
contain bugs~\cite{nargiz:dl:faults:icse2020}. Eradicating these bugs
is important to ensure continued productivity of ML-enabled
applications.


Fuzzing DL libraries is an active area of
research~\cite{Phan_ETAL_ICSE2019,AUDEE_ASE20,LEMON_FSE20,MUFFIN_ICSE22,Liu_2023,Liu_ETAL_ICSE23,Wei2022,Deng_ETAL_FSE22,Christou_ETAL_USENIX23,Deng_ETAL_ISSTA23}. Fuzzing
techniques generate input data to test 
the functions of the library --APIs,
for short-- and use some test oracle to determine the presence of a
likely bug.
One important obstacle affecting the efficiency of these tools is
\emph{the presence of input constraints}, e.g., a constraint
involving a pair of tensor\footnote{A tensor is an $n$-dimensional
array. A vector is a one-dimensional tensor whereas a matrix is a
two-dimensional tensor.} parameters of an API relating their
dimensions.
For that reason, random test generation may yield invalid inputs that
fail at runtime, reducing the efficacy of fuzzing techniques. Recent
approaches attempt to address this issue by learning constraints to
filter invalid inputs preemptively~\cite{Wei2022, Deng_ETAL_FSE22,
  docter2022,acetest2023,deepconstr2024}.  However, these methods can
be computationally intensive or limited in their capacity to precisely
infer constraints.  For instance, the constraints inferred by
ACETest~\cite{acetest2023}, a fuzzing technique that automatically
extracts constraints from source code, produces valid inputs at a rate
of only $\sim$29\%, on average.

This paper reports on a study to evaluate the effectiveness of ML
classifiers to accelerate DL library testing. 
We leverage the observation that examples of API usage are
easily accessible in this domain to train classifiers.
To obtain such examples, we generate inputs at random and, as in prior
work~\cite{acetest2023,Deng_ETAL_ISSTA23}, observe crashes
to identify positive and negative cases.\footnote{DL libraries adopt
the defensive programming practice to ``\emph{fail fast}'', i.e., to
warn developers of inputs that violate API preconditions.} We
hypothesize that using tensor shapes to encode concrete inputs and
train ML models is
(1)~\emph{accurate} to capture data relationships and determine
validity; (2)~\emph{general} as most APIs take combinations of
tensors, tuples, lists, and primitive-type data types, which can be
directly encoded in tabular form for training
classifiers~\cite{autogluon-tabular}; and (3)~\emph{efficient} to
quickly classify several inputs at once with batch inference~\cite{chetlur2014cudnn}.



Our study evaluates three aspects. \underline{First}, to
evaluate accuracy, we use the AutoGluon~\cite{autogluon-tabular}
AutoML~\cite{automl-org} framework to train and measure different
classifiers' accuracy on a set of 10,000 inputs over 98 functions of
\torch{} and 85 of \tf{},
finding that there always exists at least a classifier
achieving over 90\% accuracy.
\underline{Second}, to evaluate how general the models are, we test
the best classifier by applying it to 50,000 new input configurations
and validating its robustness in diverse testing scenarios. Our
results show that all our models generalize with over 91\% accuracy on
the 183 operators we consider. \underline{Third}, to evaluate
usefulness, we integrate the models we obtained into
ACETest~\cite{acetest2023}, a \sota{} API-level fuzzer for DL
libraries. More precisely, we evaluate if the use of ML classifiers
improve the effectiveness of ACETest by discarding the invalid inputs that
ACETest often produces (validity rate 29\%) and requesting a new
input. In this scenario, the classifier serves as a filtering
mechanism, ensuring that ACETest only executes the API on inputs the
model classifies as valid. We observe an improvement from 29.1\% to 60.7\% in
average pass rate of ACETest when incorporating the ML models.
Moreover, we show that the inclusion of the ML 
models produces a negligible impact on the 
bug finding capabilities of ACETest, 
reporting more than 90\% of the bugs when 
models are trained with at least 10\% of 
positive samples.



We make the following contributions:

\begin{itemize}[topsep=0ex,itemsep=0pt,leftmargin=1em]
\item[\Contrib{}] Idea. We propose a simple yet effective idea to
  increase the ratio of valid inputs generated per unit of time of
  API-level DL library fuzzers. The approach is particularly effective
  for APIs with complex constraints;
  
\item[\Contrib{}] A comprehensive evaluation of ML classifiers to
  check the validity of the inputs for a set of APIs from \torch\ and \tf,
  two of the most popular DL libraries today. We use
  AutoML~\cite{automl-org} to automate the discovery of such
  classifiers and find that classifiers with accuracy of at least
  \textbf{$>$90\%} exist for all APIs we analyze and they generalize
  well to previously unseen data;
  \item[\Contrib{}] A demonstration of the usefulness of ML
    classifiers by integrating them with ACETest~\cite{acetest2023}, a
    \sota\ API-level fuzzing tool of DL libraries. We show that, when
    used as a pre-filtering mechanism, to ensure that only inputs
    predicted as valid are used in the testing process, the pass rate
    of the tool, i.e., the ratio of inputs used for testing that are
    valid, can be significantly improved, reaching a pass rate of
    \textbf{$\sim$61\%}.
\end{itemize}

\vspace{1ex}\noindent{}Our artifacts are publicly available~\cite{rep-package-site}.


\section{Background and Example}
\label{background} 

\tf{}~\cite{abadi2016tensorflow} and \torch{}~\cite{paszke2019pytorch}
are two widely-used libraries that 
facilitate the training,
evaluation, and deployment of machine learning models. Developers use
a comprehensive set of APIs from those libraries to build their
models. Even though \tf\ and \torch\ are maintained by different
organizations (Google and Linux Foundation, respectively), they offer
very similar APIs.

\noindent\textbf{Tensors.}~A tensor is a multi-dimensional array. A
vector is a one-dimensional tensor. The APIs from \dl\ libraries
heavily rely on tensors for computation. For example, the API
\CodeIn{torch.bmm}\footnote{API documentation:
\url{https://pytorch.org/docs/stable/generated/torch.bmm}} performs
multiplication of two tensors.

\noindent\textbf{Input Constraints.}~APIs of DL libraries often impose
constraints on inputs restricting their usage. For instance, the API
\CodeIn{torch.bmm} expects the two input tensors to be 3-D tensors. Figure~\ref{fig:torch-bmm} shows calls to
that API with invalid and valid inputs,
respectively. Figure~\ref{fig:torch-bmm-code} shows the input
validation checks performed in the C++ backend of the \torch{}
implementation of the API \CodeIn{torch.bmm}. If some input constraint
is violated, the function \CodeIn{TORCH\_CHECK} raises an exception
that the Python front end captures and propagates to the client code
(Figure~\ref{fig:torch-bmm}) as a \CodeIn{RuntimeError}.


\begin{figure}[t!]
    \centering
    \begin{subfigure}[t]{0.48\textwidth}
      \centering
\begin{lstlisting}[language=Python,morekeywords={bmm},deletekeywords={import}]
>>> import torch
>>> # Wrong input
>>> mat1 = torch.randn(10, 3, 4)
>>> mat2 = torch.randn(10, 4)
>>> torch.bmm(mat1, mat2)
Traceback (most recent call last):
  File "<stdin>", line 1, in <module>
RuntimeError: batch2 must be a 3D tensor
>>> # Correct input
>>> mat2 = torch.randn(10, 4, 5)
>>> torch.bmm(mat1, mat2)
tensor(...)
\end{lstlisting}
\vspace{-1ex}
\caption{Negative and positive usage examples.}
\label{fig:torch-bmm}      
    \end{subfigure}%
\vspace{1ex}    
    \begin{subfigure}[t]{0.48\textwidth}
      \centering
\begin{lstlisting}[language=C++,morekeywords={TORCH\_CHECK},deletekeywords={void,const,auto}]
void common_checks_baddbmm_bmm(const Tensor& batch1, const Tensor& batch2 ...
  TORCH_CHECK(batch1.dim() == 3, "batch1 must be a 3D tensor");
  TORCH_CHECK(batch2.dim() == 3, "batch2 must be a 3D tensor");  
  const auto batch1_sizes = batch1.sizes();
  const auto batch2_sizes = batch2.sizes();  
  int64_t bs = batch1_sizes[0];
  int64_t contraction_size = batch1_sizes[2];
  int64_t res_rows = batch1_sizes[1];
  int64_t res_cols = batch2_sizes[2];
  std::vector<int64_t> output_size {bs, res_rows, res_cols};  
  TORCH_CHECK(batch2_sizes[0] == bs && batch2_sizes[1] == contraction_size, "Expected size for first two dimensions of batch2 tensor to be: [", bs, ", ", contraction_size, "] but got: [", batch2_sizes[0], ", ", batch2_sizes[1], "].");  
  ...// implementation
\end{lstlisting}
\vspace{-1ex}
\caption{Input validation of the API in the C++ backend.}
\label{fig:torch-bmm-code}      
    \end{subfigure}
    \caption{\torch{}'s \texttt{torch.bmm}. The API computes a batch matrix-matrix product
  of matrices.}
\end{figure}

\noindent\textbf{Motivation.}~Testing the APIs of \dl\ libraries is an
important problem. Prior work proposed methods to infer API input
constraints --such as the one in \CodeIn{torch.bmm}-- to accelerate
bug finding~\cite{Wei2022,
  acetest2023,deepconstr2024,docter2022} (Section~\ref{related-work}
elaborates and expands on related work).
For example, ACETest~\cite{acetest2023} uses constraint solvers to
generate inputs from input constraints inferred from the
code. Unfortunately, these techniques are inaccurate. As an example,
only 5.4\% of the inputs that ACETest generates for the \tf\ API
\CodeIn{SigmoidGrad}\footnote{\url{https://www.tensorflow.org/api_docs/python/tf/raw_ops/SigmoidGrad}}
are valid.
This API computes the gradient of the sigmoid function.


\noindent\textbf{Example.}  We illustrate the benefits of using
classification models and batch inference to accelerate fuzzing.
Considering the \CodeIn{SigmoidGrad} API,
Table~\ref{tab:comparison-with-and-without-ml} shows the breakdown of
time of ACETest and ACETest+ML across the four steps of (valid input
data) test generation: (1)~Generation, (2)~Processing, (3)~Inference,
and (4)~Execution. Column ``\emph{t}'' shows the total time in
seconds, column ``\#'' shows the number of \emph{valid inputs}
generated, and column ``\#/t'' shows the ratio of inputs generated per
second, which is our proxy of efficiency. The higher that number the
better. Note that ACETest spends the bulk of its time (31s) executing
the API. At a high level, the ML integration speeds up the test
generation process by filtering which inputs are worth executing. In
the following, we elaborate on the four steps mentioned above.


\underline{Generation} is responsible for generating inputs. The
process is identical in both approaches. For this API, the ACETest
generator produces 5K inputs in 1s.  Processing and Inference are
unique to ACETest+ML. \underline{Processing} consists of extracting
abstract values from concrete inputs (e.g., tensor data) to feed into
the model for inference. \underline{Inference} classifies inputs based
on their likelihood of being valid. ACETest+ML uses batch inference to
speed up end-to-end fuzzing time.  Batch inference utilizes optimized
memory access and hardware optimizations to enable faster inference of
\emph{multiple inputs} in one inference
query~\cite{chetlur2014cudnn,cheng2017tensorflow,Neubig_ETAL_NIPS17}.
More specifically, ACETest+ML creates a batch with the full set of 5K
abstract inputs for inference to query the model.  In this example,
inference takes about $\sim$0.1s on a batch with all of the 5K
(abstract) inputs, reducing the number of inputs to process in the
next step from 5K to 332. In contrast, ACETest without ML carries all 5K inputs to
the next stage. To emphasize the importance of batching, it is worth
noting that if ACETest+ML had queried the
model once per input (\ie, 5K times), the cost of inference would
increase to 195s, defeating the purpose of the ML integration.
Finally, the \underline{execution} step is identical in both approaches. ACETest
without ML takes 31s to produce 235 valid inputs from 5K inputs whereas ACETest+ML takes 7.4s to produce 143
valid inputs from 332 inputs.
Overall, considering the four steps, ACETest produces valid inputs at
a ratio of 7.3 inputs per second while ACETest+ML produces
valid inputs at a ratio of 14.3 inputs per second.
To sum up, considering this
scenario, \emph{we observe that the usage of classification models \textbf{and}
batch inference double the ratio of valid inputs that ACETest
generates per second}.

\begin{table}
  \footnotesize
  \centering
  \setlength{\tabcolsep}{2pt}
  \caption{\label{tab:comparison-with-and-without-ml}Runtime breakdown for 5K randomly-generated inputs for the API
    \CodeIn{raw\_ops.SigmoidGrad} with ACETest and
    ACETest+ML.} 
  \begin{tabular}{cccccccc}
    \toprule
    \multirow{2}{*}{Approach} & \multicolumn{4}{c}{Steps} &
    \multirow{2}{*}{\emph{t}} & \multirow{2}{*}{\#} &
    \multirow{2}{*}{\#/\emph{t}}\\
    \cmidrule(r){2-5}
              & Generation & Processing & \textbf{Inference} & Execution & & &
    \\
    \midrule    
    ACETest & 1s & 0s & 0s & 31s (5K) & 32s & 235 & 7.3\\
    ACETest+ML & 1s & 1.5s & 0.1s & 7.4s (332) & 10s & 143 & \cellcolor{lightgray}14.3\\
    \bottomrule
  \end{tabular}
  \vspace{-2ex}
\end{table}

\section{Study}
\label{study}

In this section we describe our empirical study 
to assess the effectiveness of
ML models in learning DL library constraints. 
We aim to answer the following research questions.

\begin{itemize}
\item[\textbf{RQ1}]How effective are ML models in learning input
  constraints of DL library APIs?
  
\item[\textbf{RQ2}]Do ML models generalize outside training data
  sets?
  
\item[\textbf{RQ3}]Do ML models improve test input generation for DL
  library APIs?
\end{itemize}

RQ1 evaluates how effective the use of ML models is to predict input
validity for DL library APIs.
To evaluate the generalizability of the best learned models, RQ2
measures their performance on data outside the training datasets.
Finally, RQ3 investigates the extent to which ML models help a
state-of-the-art fuzzing technique, namely ACETest~\cite{acetest2023},
generate valid inputs faster.

\begin{figure}[t]
\centering
\includegraphics[width=\textwidth,trim={0 14cm 15cm
    0},clip]{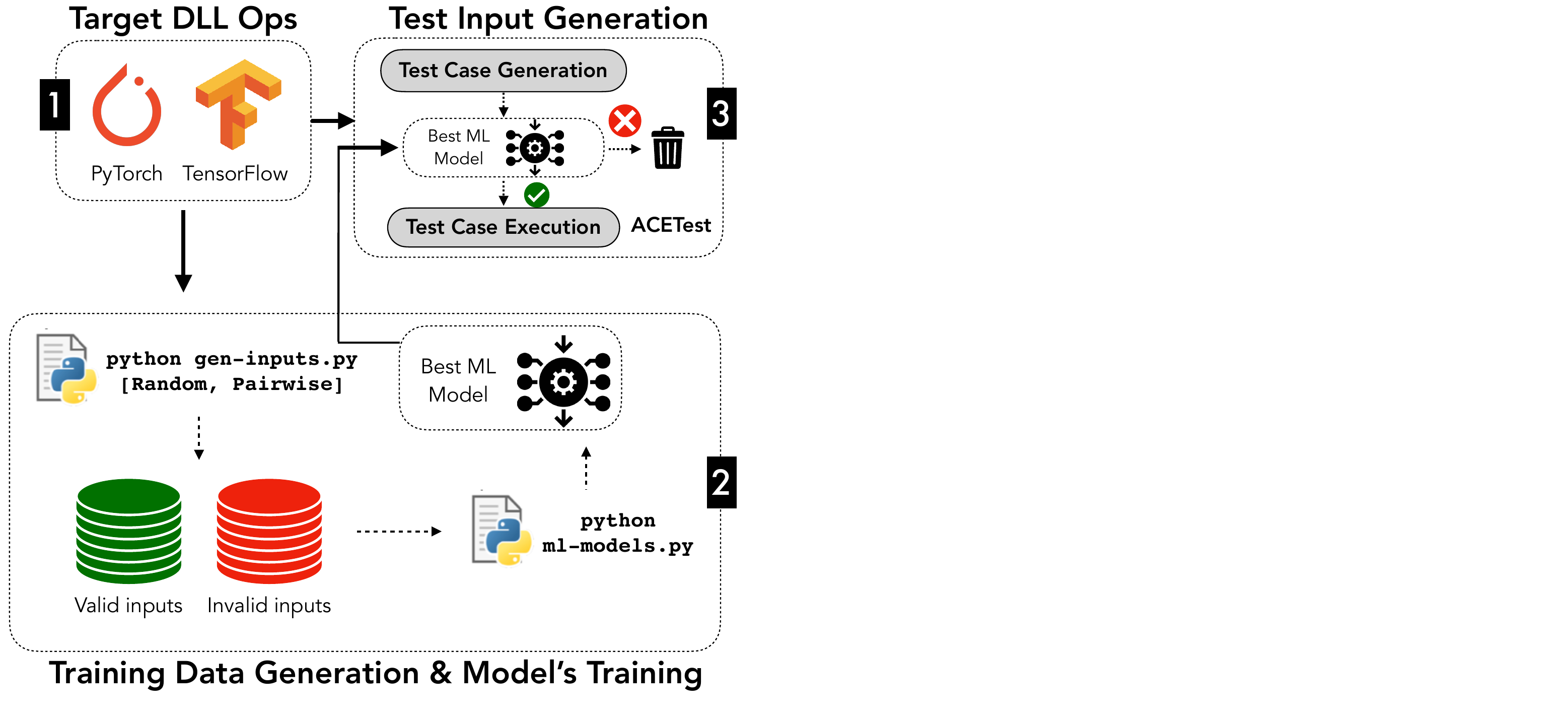}
\vspace{-3ex}
\caption{Overview of our study.}
\label{fig:study-overview}
\end{figure}

Figure~\ref{fig:study-overview} shows an overview of 
the study we conduct to answer these research questions.
We \ding{182} start from the popular DL libraries \torch{}~\cite{paszke2019pytorch} 
and \tf{}~\cite{abadi2016tensorflow},
from where we collect a dataset of DL library 
operations, covering a wide variety 
of operations and constraints.
Then, \ding{183} given a target operation to analyze,
we automatically 
generate a dataset of valid and invalid 
inputs for the operation using 
two strategies: a (i) random strategy
and a (ii) pairwise strategy.
Then, we train a family of off-the-shelf ML models 
on the automatically generated dataset 
to distinguish between valid and invalid inputs, 
and therefore capture the constraints of the operation.
We then \ding{184} assess the effectiveness and the generalizability of 
the trained models using standard metrics in ML (precision and recall).
Finally, we evaluate the models
in a practical scenario,
by integrating them into the
ACETest~\cite{acetest2023} pipeline to improve
test input generation for DL library operations. 

Below we describe how we obtained the target subjects and 
how we performed each step of our study.

\subsection{Subjects}
We use \torch{} (version 2.2.2) and \tf{} (version 2.16.2), two of the
most popular and widely used deep learning libraries today. We use a
total of 183 APIs with input constraints; 98 APIs from \torch{} and 85
APIs from \tf{}. We use the list of APIs from \freefuzz~\cite{Wei2022}
and check if executing them with random inputs would raise exception,
indicating the presence of input validation checks. The input
constraints vary in complexity.


\subsection{Training Data Generation}

To be able to train ML models to learn 
the constraints of DL library operations, 
we need to generate and label a training set 
composed of valid and invalid samples.
Given the huge input space of DL library operations,
mainly due to variability in the size and 
dimensions of tensors,
it is infeasible to generate all possible
inputs for an operation. 
Thus, we reduce the input space 
by generating training inputs within a 
certain range of values. 
For \texttt{Tensor} arguments, we limit the
maximum dimension of the tensor to 6,
and each length in the shape of the tensor
to the range [0,10]. 
For \texttt{int} arguments, we restrict the
range of values to [-100,100].
Values for \texttt{string} arguments are restricted 
to a set of predefined strings, obtained from the documentation of the
corresponding API. 
Finally, arguments of type \texttt{float} and
\texttt{bool} have no restrictions.
Additionally, we do not impose any restriction 
on the tensor elements. 

We defined the ranges of values for each
argument type based on common values used 
in practice.
To obtain these values, we collected code snippets from 
\torch{} issue reports created in the period between
Sep 14, 2021 - Jun 27, 2024.
Then, we leverage the Llama3\footnote{\url{https://github.com/meta-llama/llama3}}
Large Language Model 
to analyze the code snippets and 
automatically extract tensor values and shapes.
We do this by providing a structured prompt
to the model in order to guide it to produce a
standardized JSON-like output 
for any recognized tensor in the issue report. 
Through this process, we collected 910 tensor 
configurations, where 
the maximum number of dimensions was 6.
We limited the maximum size of a dimension to 10, which is the largest size
among the top-5 most frequently occurring shapes.

Given a target API, using the ranges of values defined above,
we create a training set comprised of 10K samples using
two strategies:

\textbf{Random}. In the Random strategy, 
for each argument of the target operation,
we randomly generate 10K tuples, 
where each element is of the corresponding argument type.
Then, 
we execute the target operation on each input tuple.
If the operation
raises an exception, we consider
the input as invalid, as it 
is rejected by the input validation code. Otherwise, if the operation 
returns successfully, we consider the
input to be valid.
Note that the assumption that the existing 
input validation code is correct is a common 
assumption in the literature on
constraint generation~\cite{acetest2023,deepconstr2024}.

\textbf{Pairwise}. In the Pairwise strategy,
we use pairwise combinatorial
testing~\cite{bach2004pairwise,Amman_Offutt_2016} to generate
inputs. Pairwise testing examines every possible combination of values
for every pair of input parameters. Intuitively, this strategy samples
the input space more uniformly compared to the Random strategy.  
If all pairwise combinations are covered before reaching the 10K
inputs we aim to generate, we start over from the first combination
until reaching 10K. It is worth noting that string and primitive-type
values are generated at random in this process. Similar to the Random
strategy, we determine the label of each input by executing the target
API.

\subsection{ML Model Training}

For a target API,
we train ML models using the corresponding
training set generated in the previous step.
The generated set of 10K samples
is split into a training set and a test set using a 80:20 ratio.
To account for the randomness,
we repeat the training process, 
including the generation of the dataset, 
10 times for each target API, 
using a different random seed each time.

Features used to train the models
are essentially the raw input values, 
encoded depending on their type.
For \texttt{int} and \texttt{float} arguments,
we use their values directly.
For \texttt{string} arguments, we encode
them using a mapping to integer values.
For \texttt{Tensor} arguments, we discard 
the actual tensor elements, and
encode just the shape of the tensor, which is a 
tuple of integers. 
Finally, for \texttt{bool} arguments, we 
encode them as 0 and 1.
This encoding is necessary to allow the
ML models to properly process the input values.

To actually train different ML models we rely on the 
AutoGluon~\cite{autogluon-site}
Python library, which automates ML tasks.
During training, AutoGluon fits a family of 
various ML models, ranging from 
off-the-shelf boosted trees to 
customized neural networks.
After training, a leaderboard of the best performing
models can be consulted, and the user can
select the best performing
models for further evaluation.
Below we list the models which more 
frequently appear as the best performing
models in our experiments (the complete list 
can be found online in the AutoGluon documentation).

\noindent
\textbf{CatBoost}~\cite{catboost2018} is an ML algorithm 
based on \emph{gradient boosting} on decision trees.
Gradient boosting is an ML ensemble technique 
that combines predictions from multiple 
weak models, typically decision trees,
in a sequential manner, so that each 
new model corrects the errors of its predecessor.
\textbf{LightGBM}~\cite{lightgbm2017}, 
also based on gradient boosting,
is characterized by a histogram-based learning approach, 
which constructs histograms of continuous features, subsequently utilizing 
these discrete bins to find the optimal split over features. 
\textbf{XGBoost}~\cite{xgboost2016}, or 
Extreme Gradient Boosting,
is an optimized version of 
gradient boosting. 
One of the biggest strengths of 
XGBoost is its speed and efficiency. 
\textbf{NeuralNetFastAI} is a model from
FastAI~\cite{fastai-site},
a library that provides fast 
neural network training.
\textbf{ExtraTrees} is an ensemble ML model, available in the
scikit-learn library~\cite{scikitlearn-site}. It
trains numerous decision trees and aggregates the results from the
group of decision trees to output a prediction.

It is important to note that 
AutoGluon uses cross-validation
when training the available ML models.
Cross-validation is a technique 
that subsequently 
splits the training data into $k$ folds,
trains the model on $k-1$ folds, and
evaluates it on the remaining fold.
This process is repeated $k$ times,
and it is used with the goal of 
reducing over-fitting and increasing
confidence in the model's
performance.

\subsection{ML Models Evaluation}

To evaluate the performance of the ML models we 
use the standard metrics
\emph{precision} and \emph{recall}.
These metrics are computed from the
model predictions on the test set,
which can be classified into 
true positives, false positives, true negatives, and false negatives.

\noindent
\textbf{True Positives (TP).} A true positive is 
an input that is accepted by the input validation of the
target operation and the model correctly
predicts as valid.

\noindent
\textbf{False Positives (FP).} A false positive 
is an input that is rejected by the input validation of the 
operation, but the model incorrectly
predicts as valid. In a practical 
scenario in which the model is used to filter inputs, 
a false positive would lead to the
model accepting an invalid input, and therefore 
needlessly executing the operation.

\noindent
\textbf{True Negatives (TN).} A true negative is
an input that is rejected by the input validation of the
target operation and the model correctly
predicts as invalid.

\noindent
\textbf{False Negatives (FN).} A false negative is an input that is
accepted by the input validation of the target operation, but the
model incorrectly predicts as invalid. A
false negative indicates that a model would reject a valid,
potentially bug-revealing, input.

From these four prediction types, we measure precision and recall of a
classification model as follows: \emph{Precision=TP/(TP+FP)},
\emph{Recall=TP/(TP+FN)}. Essentially, precision measures the
proportion of all the inputs predicted as valid that are actually
valid; it is higher when the model makes fewer false positive
predictions. Recall, on the other hand, measures the proportion of
all valid inputs that were correctly predicted as valid; it is higher
when the model makes fewer false negative predictions.


\subsection{Test Input Generation}

In the last part of our study, 
we focus on the practical application
of the trained ML models.
Concretely, we study how the 
obtained classifiers can be used 
to improve test input generation
for DL library operations.
To this end, given a target operation,
we integrate our classifiers 
into the ACETest test generation
pipeline. 

ACETest~\cite{acetest2023} generates test inputs for DL library 
operations by generating solutions 
to its previously extracted constraints.
Since these constraints are expressed 
as Z3 formulas, ACETest uses the Z3 solver
to generate the test inputs.
In our integration, before actually 
invoking the target operation on the
generated inputs, we first pass them
through the best performing ML model for 
the operation.
The classifier acts as a pre-filter, checking input validity
before executing the operation.
If the model predicts the input as valid,
we proceed with the execution of the operation.
Otherwise, we discard the input and
generate a new one until we process 
all the inputs that ACETest was instructed to generate.
In our experiments, we set the number of inputs
to generate to 5,000.

Following the ACETest evaluation, 
we assess the improvement achieved by incorporating the ML 
classifiers using the 
\emph{pass rate} metric. 
The pass rate evaluates the ratio of 
generated test cases that can pass all input
validation checks of the target operation.
Either for the standard ACETest, 
or for the ACETest with the integrated
ML model, we can measure the pass rate 
by computing the ratio of inputs that 
do not trigger an exception when
executed on the target operation.
Additionally, in both cases, we measure 
the time taken for the 
whole testing process as well as 
the ratio of valid inputs 
generated per second.
Finally, we assess how the inclusion 
of the ML models affects the
bug finding capabilities of ACETest.

\subsection{Implementation and Setup}

All the experiments we performed are implemented 
as Python scripts, using \torch{} 2.2.2 and 
\tf{} 2.16.2.
To train the ML models, we rely on AutoGluon 1.1.1. 
For evaluating the improvement on test input generation, 
we extend the ACETest tool publicly available 
on GitHub~\cite{acetest-tool},
with the ability to 
predict validity of the generated inputs
using our trained ML classifiers.

We run all our experiments on a workstation 
with a Xeon Gold 6154 CPU (3GHz),
with 128 GB of RAM, running Debian/GNU Linux 11. 
Finally, all the scripts and data required 
to obtain the results presented in this paper
are available online~\cite{rep-package-site}.

\section{Experimental Results}
\label{results}

This section presents the results for 
each research question.

\subsection{Effectiveness of ML models in learning DL library constraints (RQ1)}

\begin{table}[t!]
  \caption{\label{rq1-effectiveness}ML models effectiveness in learning DL library constraints, reporting averages over 10 runs for positive and negative samples, generation time, and best-model precision and recall per data generation strategy.}
  \begin{center}
  \begin{scriptsize}
  \renewcommand{\arraystretch}{1.2} 
  \setlength{\tabcolsep}{0.4em}
  \scalebox{1.0}{
  \begin{tabular}{lr|lrrr|rr}
  \toprule
  \multicolumn{2}{c}{\textbf{Target}} &   \multicolumn{4}{c}{\textbf{Training Data}} & \multicolumn{2}{c}{\textbf{Model Performance}} \\
  \textbf{DL Library} & \textbf{\#APIs} & \textbf{Technique} & \textbf{\#Pos.} & \textbf{\#Neg.} & \textbf{Time (sec.)} & \textbf{Precision} & \textbf{Recall} \\
  \midrule
  \torch{} & 98 & RANDOM & 1,633 & 8,291 & 46.87 & 87\% & 79\% \\
  & & PAIRWISE & 1,633 & 8,263 & 49.77 & 88\% & 82\% \\
  \midrule
  \tf{} & 85 & RANDOM & 1,699 & 8,300 & 5.18 & 90\% & 78\% \\
  & & PAIRWISE & 1,529 & 8,470 & 5.71 & 91\% & 80\% \\
  \bottomrule
  \end{tabular}}
  \end{scriptsize}
  \end{center}
  \end{table}

\begin{figure*}[t]
  \begin{minipage}{.48\textwidth}
    \centering
    \includegraphics[width=\textwidth]{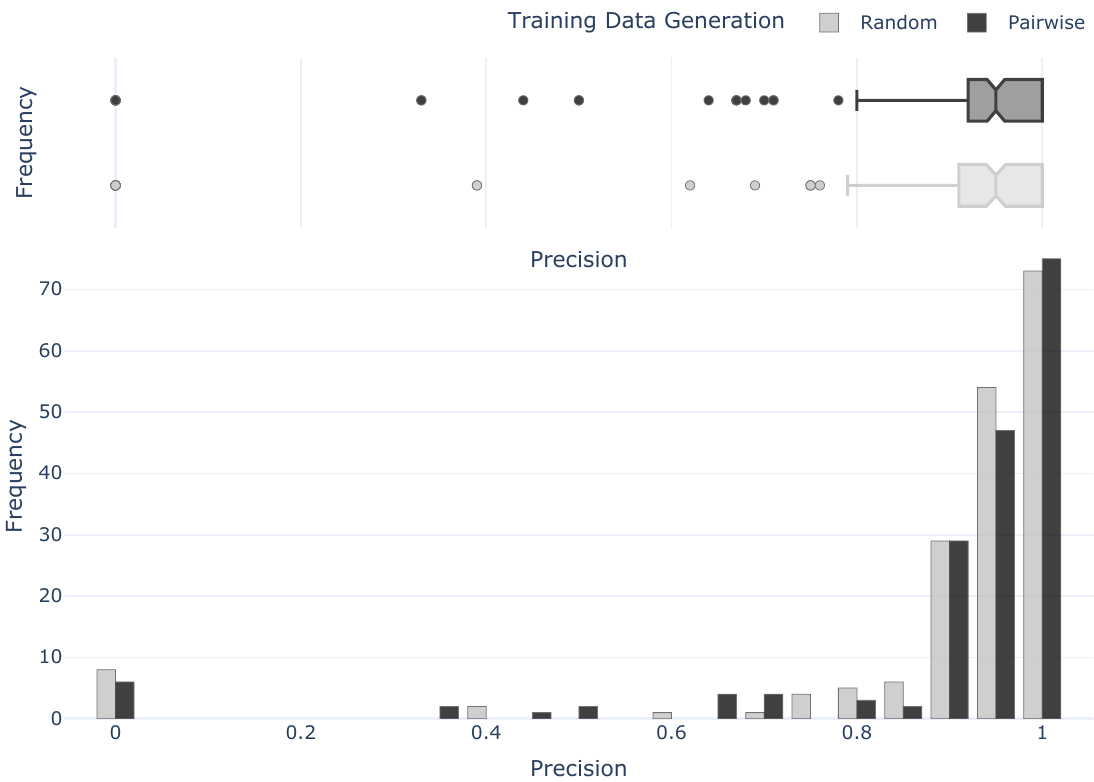}
    \label{fig:all-precision-rq1}
  \end{minipage}
  \hfill
  \begin{minipage}{.48\textwidth}
    \centering
    \includegraphics[width=\textwidth]{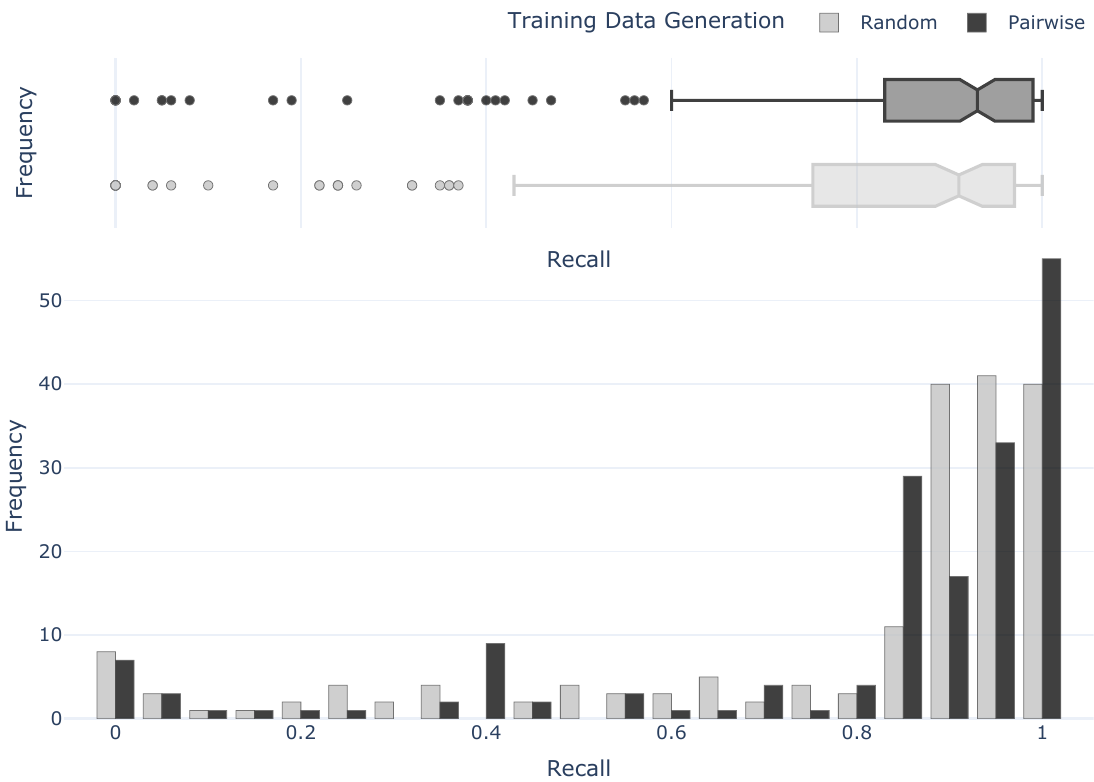}
    \label{fig:all-recall-rq1}
  \end{minipage}
  \vspace{-3ex}
  \caption{Distribution of the precision and recall values achieved by ML models when learning DL library operation constraints. Each plot reports the frequency of precision/recall values for the random (left) and pairwise (right) training data generation strategies.}
  \label{fig:all-precision-recall-rq1}
\end{figure*}

Table~\ref{rq1-effectiveness} shows the results of our study for RQ1. 
For each 
training data generation strategy (random and pairwise), 
we report the average number of positive and 
negative samples in the training set,
the average time taken to generate the
training set. Additionally, we report 
the corresponding average precision
and recall values achieved by 
the best performing models.

We observe that, on average, ML models can achieve 
high precision and recall in learning 
the properties of DL library constraints, 
with up to 88\% precision and 82\% recall
in the case of \torch{} operations, 
and up to 91\% precision and 80\% recall 
in the case of \tf{} operations.
Moreover, we observe that using the 
pairwise strategy consistently leads to 
better precision and recall
values in both DL libraries,
compared to the random strategy.

Notably, these results show that 
state-of-the-art ML models can achieve 
impressive performance in classifying 
the input validity for DL library operations, 
even with a relatively small
percentage of positive samples in the training set. 
For each target API, our training data strategies  
generate (on average) up to $\sim$17\% 
of positive samples.
Though this low percentage of positive samples
may pose over-fitting risks, as we show in 
Section~\ref{rq2-results}, the models
are able to generalize well to unseen data.

It is worth mentioning that 
the time taken to generate the training data is
not significantly different between the two strategies, 
with an increase of $\sim$3 seconds
for the pairwise strategy just in the case of \torch{} operations.
Moreover, it takes less than a minute
to generate the training data on average for each target API.

\subsubsection{Random vs Pairwise.}

Let's now compare 
in more detail the performance of the models
trained with each strategy.
Figure~\ref{fig:all-precision-recall-rq1} 
shows the distribution of 
precision and recall values achieved by the
ML models considering the two training data generation
strategies and all the analyzed DL library operations.  
Overall, the pairwise strategy leads to an 
improvement in the performance of the
models over the random strategy.
Though on average there is no significant
difference between the two strategies, 
we believe the diversity of samples generated by 
the pairwise strategy is the key factor
contributing to the higher 
precision and recall values.

Considering 80\% as the threshold for a good model,
the random strategy allows to achieve a precision of at least 
80\% in 89\% of the target operations, 
and a recall above 
80\% in 73\% of the target operations.
When we use the pairwise strategy,
the amount of cases for which we obtain a precision
of at least 80\% remains the same. 
However, the amount of cases for which 
we obtain a recall
above 80\% increases to 79\%, respectively.
Higher recall values are preferable as they mean fewer false negatives,
so fewer valid inputs are incorrectly discarded.
This suggests that the pairwise strategy is more effective for testing.

In both strategies, for the rest of the cases 
the precision and recall values are equally 
distributed between 0\% and 80\%, 
with a slight tendency to 0\%. Below we 
discuss in more detail operations 
in which the models achieve an outstanding 
performance 
and operations in which the performance is poor.

\subsubsection{Best Performing Models.}

There are various operations with complex constraints 
for which the models achieve a remarkable performance.
Table~\ref{rq1-example-complex-constr} shows some examples
of operations with complex constraints.
For instance, from \torch{}, the operation 
\CodeIn{broadcast\_to(input, shape)}
requires that the length of the shape
is greater or equal to the dimension of the input tensor;
the \CodeIn{cartesian\_prod(*tensors)} operation requires
all tensors must be 1D tensors;
and the \CodeIn{MaxPool2d} operation needs 
the kernel size to be defined in terms of the 
input size and padding, while the stride and 
padding also have to satisfy certain conditions.
In the case of \tf{}, some example operations 
for which the models achieve a good performance are 
the \CodeIn{MatrixInverse(tensor)} operation, which requires
a tensor whose inner-most 2 dimensions form square matrices;
the \CodeIn{tf.math.top\_k(input,k)} operation, in which 
the tensor can have any dimensions but the last one 
must be at least \CodeIn{k}; 
and the \CodeIn{split(value, num\_splits, axis, num)} operation,
where \CodeIn{num\_splits} value must evenly divide 
the value \CodeIn{value.shape[axis]} and \CodeIn{axis} must respect 
a range related to the tensor dimension.

\begin{table}
  \caption{\label{best-models-table}Top-5 best performing models for API constraint learning.}
  \vspace{-2ex}
  \begin{center}
  \begin{scriptsize}
  \renewcommand{\arraystretch}{1} 
  \setlength{\tabcolsep}{0.4em}
  \scalebox{1.3}{
  \begin{tabular}{lr|rr}
  \toprule
  \textbf{Model} & \textbf{Frequency} & \textbf{Precision (avg)} & \textbf{Recall (avg)} \\
  \midrule
  CatBoost & 47.1\% & 94\% & 86\% \\
  LightGBM & 18.8\% & 95\% & 83\% \\
  NeuralNetFastAI & 12\% & 81\% & 62\% \\
  XGBoost & 11.2\% & 87\% & 82\% \\
  ExtraTrees & 10.9\% & 93\% & 88\% \\
  \bottomrule
  \end{tabular}
  }
  \end{scriptsize}
  \end{center}
\end{table}

\begin{table}[t!]
  \scriptsize
  \caption{\label{rq1-example-complex-constr} Operations with complex constraints 
    for which the trained ML models achieve a remarkable performance.}
  \vspace{-2ex}
  \setlength{\tabcolsep}{1pt}
  \centering
  \subfloat[\torch]{
  \scalebox{0.9}{
    \begin{tabular}{lllrr}
      \toprule
      \multicolumn{1}{c}{\textbf{API}} & \multicolumn{1}{c}{\textbf{Constraint}} & \multicolumn{3}{c}{\textbf{Performance}} \\
      & & \textbf{Model} & \textbf{Precision} & \textbf{Recall} \\
      \midrule
      \texttt{broadcast\_to(input, shape)} & \texttt{len(shape) >= input.dim()} & LightGBM & 90\% & 90\% \\
      \texttt{cartesian\_prod(*tensors)} & \texttt{all t in tensors: t.dim() = 1} & CatBoost & 83\% & 100\% \\
      \texttt{MaxPool2d(input, kernel\_size,} & \texttt{kernel\_size} $\le$ \texttt{input.size() + 2} $\times$ & CatBoost & 100\% & 97\% \\
      \texttt{    stride, padding)}& \texttt{padding} $\land$ \texttt{stride > 0} $\land$ \texttt{padding >= 0} & & & \\
      \bottomrule
  \end{tabular}}
  }\vspace{1ex}
  \subfloat[\tf]{
    \scalebox{0.9}{
      \begin{tabular}{lllrr}
        \toprule
        \texttt{MatrixInverse(tensor)} & \texttt{tensor.shape() = [..., M, M]} & LightGBM & 100\% & 95\% \\
        \texttt{math.top\_k(input,k)} & \texttt{tensor.shape() = [..., N] $\land$ \texttt{N} $\ge$ \texttt{k}} & ExtraTrees & 100\% & 100\% \\
        \texttt{split(value, num\_splits, axis,} & \texttt{value.shape()[axis]} // \texttt{num\_splits = 0} & LightGBM & 97\% & 84\% \\
        \texttt{  num)}& \texttt{axis} $\in$ \texttt{[-value.dim(),value.dim()]} &  & &  \\
        \bottomrule
    \end{tabular}}
  }
\end{table}

As Table~\ref{rq1-example-complex-constr} shows, 
the trained ML models are capable of capturing 
these complex constraints with high accuracy.
In Table~\ref{best-models-table} we show the 
best performing models considering 
all the target APIs and all training runs, 
the frequency of each model as the best performing model,
and their average precision and recall values.
Notably, the gradient boosting models 
(\texttt{CatBoost}, \texttt{LightGBM}, and \texttt{XGBoost})
are the most frequent best performing models
in $\sim$77\% of the cases, 
with \texttt{CatBoost} leading
in $\sim$47\% of the cases.
In the remaining cases, the \texttt{NeuralNetFastAI} 
and \texttt{ExtraTrees} models are the best performing
models, with a frequency of 12\% 
and $\sim$11\%, respectively.


However, for some operations the ML models 
achieve a poor performance.
Some examples of these cases are the operations 
\CodeIn{addcmul}, \CodeIn{bmm} or \CodeIn{einsum}
from \torch{}, 
where the precision and recall values 
are nearly 0\%.
The poor performance is mainly due to the training 
data generation process not producing enough positive samples.
In these cases, our strategies 
generate less than 52 positive samples.
For instance, the operation \CodeIn{bmm(input, mat2)}
requires both input tensors to be 3D tensors 
and contain the same number of matrices, e.g., 
\CodeIn{input = (b, n, m)} and \CodeIn{mat2 = (b, m, p)}.
In this case, our training data generation process
generates only 4 positive samples on average,
which results in an imbalanced training set.


\subsection{Generalization of ML models (RQ2)}
\label{rq2-results}

\begin{figure}[t]
  \begin{minipage}{.45\textwidth}
    \centering
    \includegraphics[width=\textwidth]{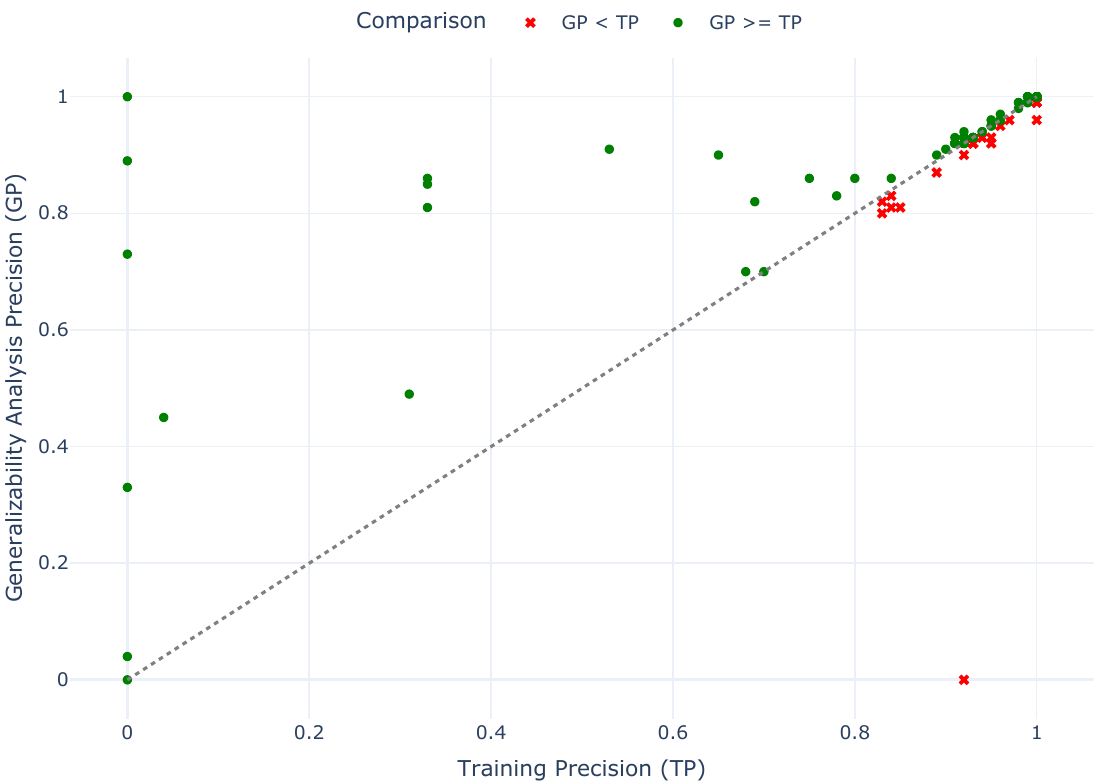}
    \label{fig:all-precision-rq2}
  \end{minipage}
  \hfill
  \begin{minipage}{.45\textwidth}
  \centering
  \includegraphics[width=\textwidth]{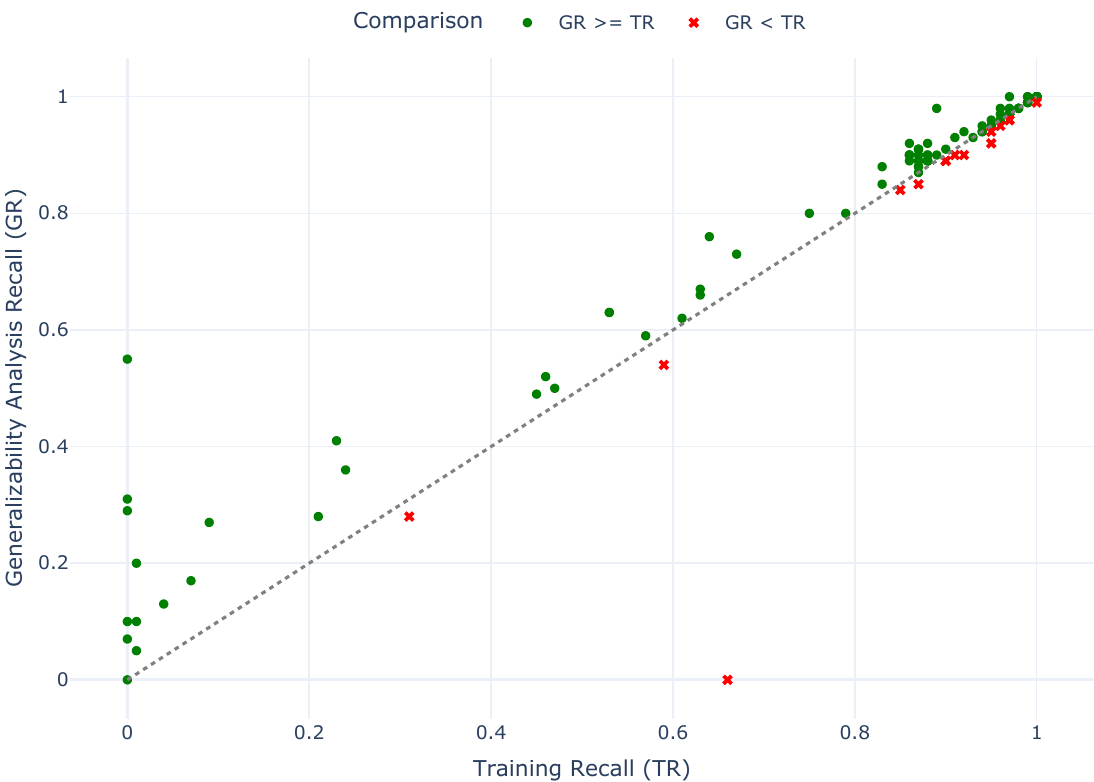}
  \label{fig:all-recall-rq2}
  \end{minipage}
  \caption{Comparison of Precision/Recall obtained during training the
    ML models and during their evaluation on a new dataset of 50,000
    samples.
    Each dot represents, for each precision (recall) value achieved during training, 
the corresponding precision (recall) value achieved
during the generalization analysis.
Green circles indicate a better precision (recall) value during generalization,
while red crosses indicate a worse precision (recall) value during generalization.
}
  \label{fig:all-precision-recall-rq2}
\end{figure}

It is well known that ML models can achieve a high performance
on the training set, but fail to generalize to unseen data.
To understand the generalization capabilities in our study,
we analyze the best performing models on a new dataset
of 50,000 samples, and measure the precision and recall
achieved on this new dataset.
Figure~\ref{fig:all-precision-recall-rq2} shows the comparison
of precision and recall values achieved during training
and during the evaluation on the new dataset.

Overall, the models are able to generalize
well to unseen data. 
In only 20\% of the cases, precision 
is lower during generalization.
However, most of these cases are above the 80\% threshold
and very close to the diagonal line, indicating a small difference
between the precision values achieved during training and generalization.
In the case of recall, only in 12\% of the cases the model
achieved a lower recall value during the generalization
analysis. Again, 
most of these
cases are very close to the diagonal line, indicating
a small difference between the recall values.

There is one outlier case in both metrics, 
corresponding to the \CodeIn{torch.dot(input, tensor)} operation.
For this operation, 
the model achieved
a precision of 92\% during training, which drastically 
decreased to 0\% during the generalization analysis.
That is, though 181 positive samples were generated
among the 50,000, the model was not able to
correctly predict any of them as valid, 
resulting in 0 true positives.
Similarly, the recall value decreased from 66\% to 0\%.
This operation requires the two 
input tensors to 
be 1D tensors and have the same number of elements. 
Though the model performs well during training, 
only an average of 37 positive samples is generated
for this operation, which may be leading 
to over-fitting.

\subsection{Test Input Generation improvement with ML models (RQ3)}

  \begin{table}[t!]
    \caption{\label{rq3-test-generation}
      ML-enhanced test input generation for DL library operations. For each target API, 5,000 inputs are generated and the ACETest pipeline is executed with and without ML models. We report the average testing time, total analysis time, average number of invalid inputs, average pass rate, and average number of valid inputs generated per second.}
    \vspace{-2ex}
    \begin{center}
    \begin{scriptsize}
    \scalebox{1.1}{
    \begin{tabular}{l|rr|rrr}
    \toprule
    \textbf{Approach} & \multicolumn{2}{c|}{\textbf{Analysis Time}} & \multicolumn{3}{c}{\textbf{API Inputs}}  \\
    & \textbf{Avg.} & \textbf{Total} & \textbf{\#Invalid} & \textbf{Pass Rate} & \textbf{\#Valid/s}  \\
    \midrule
    \rowcolor{gray!20} \multicolumn{6}{c}{All APIs (41)} \\  
    ACETest & 54.8s & 2,245s & 3,542.9 & 29.1\% & 58.3 \\
    ACETest+ML & 21.4s & 876s & 354 & 60.7\% & 42 \\
    \midrule
    \rowcolor{gray!20} \multicolumn{6}{c}{ACETest Pass Rate >= 40\% (10)} \\
    ACETest & 42.5s & 425s & 846.5 & 83.1\% & 197.7 \\
    ACETest+ML & 35.4s & 354s & 305.4 & 90.9\% & 116.9 \\
    \midrule
    \rowcolor{gray!20} \multicolumn{6}{c}{ACETest Pass Rate < 40\% (31)} \\
    ACETest & 58.7s & 1,820s & 4,412.7 & 11.8\% & 13.4 \\
    ACETest+ML & 16.9s & 522s & 369.7 & 51\% & 17.9 \\
    \bottomrule
    \end{tabular}}
    \end{scriptsize}
    \end{center}
    \end{table}

  Table~\ref{rq3-test-generation} shows the results
  of our experiments for RQ3.
  We report the performance 
  of the test input generation process for two approaches. 
  ACETest refers to a \sota{} test input generation 
  technique~\cite{acetest2023}, 
  while ACETest+ML is the
  extension of the ACETest pipeline that incorporates, 
  for a target operation,
  the best performing ML model in order to 
  predict input validity before actually executing
  the operation.
  For each technique and DL library, 
  we report the average time
  of the test input generation process 
  and the total analysis time.
  We also report the average number 
  of invalid inputs generated,
  with the average pass rate 
  achieved by the testing process.
  Furthermore, we report the average 
  number of valid inputs generated per second, 
  our proxy for the efficiency 
  of the testing process (Section~\ref{background}).
  These metrics are reported for three 
  different groups of target operations:
  all the analyzed APIs,
  the APIs for which ACETest achieves a pass rate
  of at least 40\% (easy), and the APIs for which ACETest
  achieves a pass rate below 40\% (hard).

  Note that, for this analysis, we only consider 
  the APIs that are both part of 
  our dataset as well as  
  the dataset of the ACETest tool. 
  This is a total of 41 APIs considering 
  both libraries.

  \subsubsection{Pass Rate (Accuracy)}

  Notably, considering all the analyzed functions
  from \torch{} and \tf{}, 
  the average pass rate goes from 29.1\% in 
  the standard ACETest process to 60.7\%
  when incorporating the ML models,
  representing an improvement of 108.5\%.
  Moreover, even when the pass rate of ACETest 
  is relatively high ($>$40\%), 
  the ML models are able to increase the average 
  pass rate from 83.1\% to 90.9\%.
  When ACETest achieves a pass rate below 40\%,
  the improvement is even more significant,
  from 11.8\% to 51\%.
  
  The main reason behind this improvement is that
  the ML models are able to correctly 
  discard many of the invalid inputs generated
  by the ACETest process. 
  For instance, considering all the operations,
  the average number of invalid inputs generated 
  decreases, on average, from 3,542.9 to 354.
  The pass rate improvement, 
  as well as the considerable reduction in the
  amount of invalid inputs used 
  to unnecessarily
  test the target operation,
  indicates that the ML models can be effectively
  used to improve the test input generation process
  for DL library operations.

  \begin{figure}[t]
    \centering
    \includegraphics[width=\columnwidth]{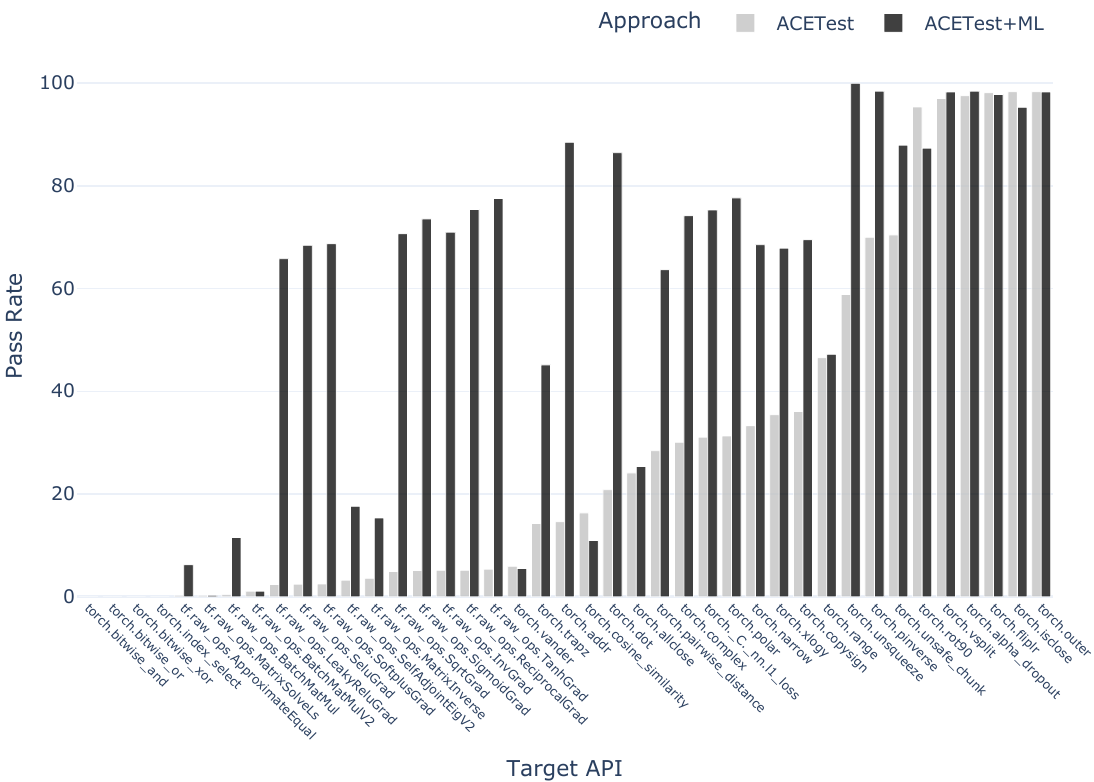}
    \vspace{-3ex}
    \caption{Pass Rate improvement of ACETest+ML.}
    \label{fig:acetest-ml-rq3}
  \end{figure}

  To better analyze the improvement achieved
  by incorporating the ML models in the ACETest pipeline,
  we analyze the performance of the test input generation
  process for each target operation.
  Figure~\ref{fig:acetest-ml-rq3} shows the improvement
  achieved on each target function.
  For each target, 
  we plot the pass rate achieved by the standard ACETest
  process (ACETest) and the pass rate achieved by
  ACETest extended with the pre-filtering mechanism
  using the best performing ML model (ACETest+ML).
  The functions are shown in increasing order of the pass rate
  achieved by ACETest.

  In some operations, the improvement is remarkable.
  For instance, for \CodeIn{SeluGrad} from \tf{},
  a common activation function 
  computing the gradients for the 
  scaled exponential linear (Selu) operation,
  the pass rate goes from 2.4\% to 68.4\%.
  In the case of ACETest, 4,876 invalid inputs are generated
  out of the 5,000 inputs produced to test the function;
  while in the case of ACETest+ML, only 168 
  inputs were used to test the function with 53 of 
  them being invalid. 
  There are various other cases with a considerable increase 
  in the pass rate, such as \CodeIn{torch.addr} and 
  \CodeIn{torch.pairwise\_distance}.
  These examples 
  illustrate the ability of the ML classifiers to 
  discard most of the actually invalid inputs produced 
  by the ACETest process.

  There are, however, some operations for which the 
  pass rate is 0 in both approaches. 
  Some examples of these cases are the 
  \CodeIn{torch.index\_select}, 
  \CodeIn{torch.bitwise\_and} and \CodeIn{torch.bitwise\_or} operations 
  from \torch{}.
  Although most of the constraints generated by ACETest are sound,
  the inferred constraints for these specific operations are too 
  weak, which leads to the generation of many invalid inputs.
  For instance, for the three mentioned operations the 5,000 inputs generated by ACETest are all invalid.
  This also affects the performance of ACETest+ML,
  as there are no valid inputs to test the operations.
  Nevertheless, it is worth remarking 
  that for these cases the ML models 
  correctly predict all the inputs as invalid,
  preventing unnecessarily testing the operations.


  \begin{figure}[t]
    \centering
    \includegraphics[width=\columnwidth]{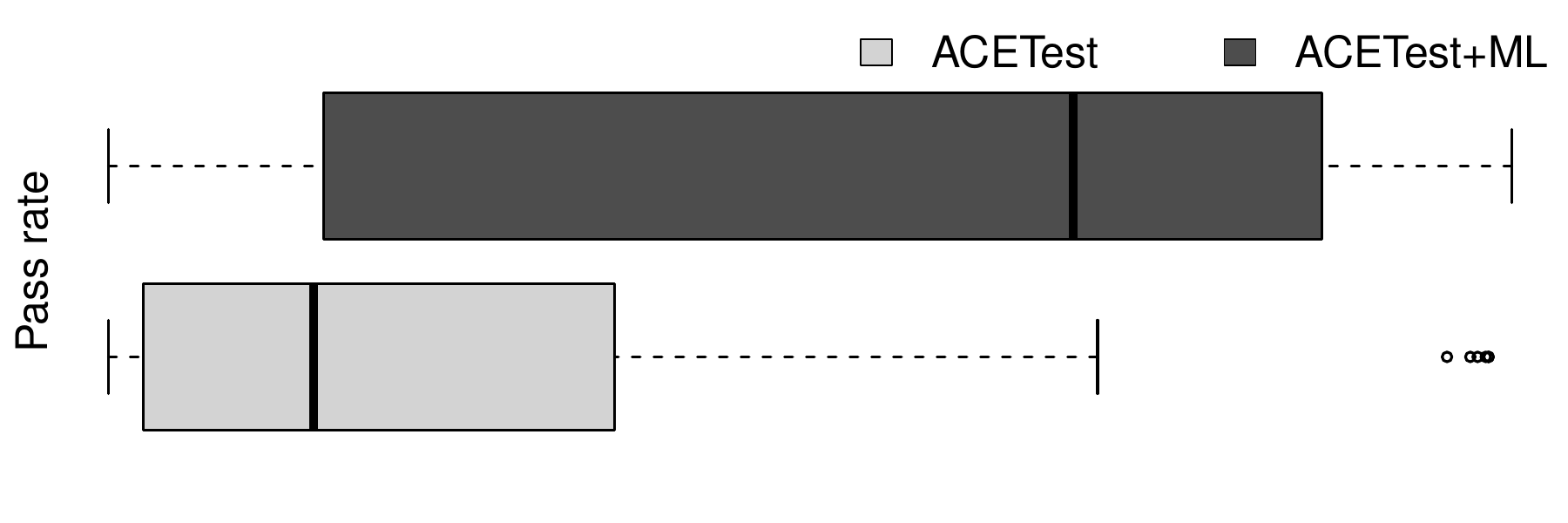}
    \vspace{-3ex}
    \caption{Pass rate distribution for ACETest and ACETest+ML.}
    \label{fig:pass-rate-rq3}
  \end{figure}

  To analyze the statistical significance of the difference between
  the pass rates achieved by ACETest and ACETest+ML after running each
  technique on \nAPIpassrate{} APIs, we first run the
  Kolmogorov-Smirnov test \cite{massey1951kolmogorov} on the two
  distributions to test the normality of the distributions. Both
  distributions produce $p$-values less than 0.05
  (\ksPvalueACETestpassrate{} for ACETest and
  \ksPvalueACETestwMLpassrate{} for ACETest+ML) indicating the
  non-normality of the distributions. This result informs us to use
  the nonparametric Wilcoxon Rank Sum
  test~\cite{wilcoxon1992individual} to calculate statistical
  significance. From this test, we obtain a $p$-value of
  \wilcoxonPvaluepassrate{}, which rejects the null hypothesis that
  the difference between the distributions (of pass rates from ACETest
  and ACETest+ML) do not differ with statistical significance.  For
  reference, a $p$-value below 0.05 is sufficient to reject the null
  hypothesis. Furthermore, we measure the Cohen's $d$
  \cite{rosenthal1994parametric} value between the distributions
  ACETest+ML and ACETest to calculate the effect size, i.e., the $d$
  value measures the magnitude of the difference between a pair of
  distributions. We find a Cohen's d value of \cohensDvaluepassrate{}
  indicating that the ML-based pre-filtering to have a
  \effectSizepassrate{} effect size. Figure~\ref{fig:pass-rate-rq3}
  shows the distribution of pass rates.

  \subsubsection{Analysis Time and Valid Inputs per Second (Efficiency)}
  Including the ML models in the ACETest pipeline
  results in a significant improvement on 
  the average time taken 
  to generate the inputs.
  Considering all the analyzed APIs 
  from \torch{} and \tf{},
  the average time taken to generate the inputs
  decreases from 54.8s to 21.4s,
  while the total analysis time decreases 
  from 2,245s to 876s.
  
  Though the time is considerably reduced, 
  many valid inputs could be generated
  and incorrectly discarded by the ML models.
  To more accurately assess the 
  efficiency of the test input generation process,
  we report the average number of valid inputs generated per second.
  Considering all the APIs, the average number of valid inputs generated
  per second is better in the case of ACETest (58.3) than in the case of ACETest+ML (42). 
  This is also the case for the APIs with a pass rate above 40\%.
  However, for APIs for which ACETest performs poorly 
  (pass rate below 40\%), the average number of valid inputs generated per second is better in the case of ACETest+ML (17.9) than in the case of ACETest (13.4).
  Moreover, if we only consider the operations 
  for which the corresponding training data 
  contains a reasonable proportion of valid inputs (more than 10\%),
  the metric increases from 20.1 to 39.3 valid inputs per second.

  It is important to remark 
  that our batching 
  inference plays a crucial role in the efficiency of our approach.
  We note that, on average, 
  enabling batching
  reduces the analysis time 
  by $\sim$49\%.
  These results show that incorporating ML models
  in the ACETest pipeline can significantly reduce the 
  time needed to generate the inputs, and, 
  when the ACETest process has a low pass rate, 
  the ML models can help to generate valid inputs 
  more efficiently.

  \subsubsection{Bug Detection}
  
  \begin{figure}[t]
    \centering
    \includegraphics[width=\columnwidth]{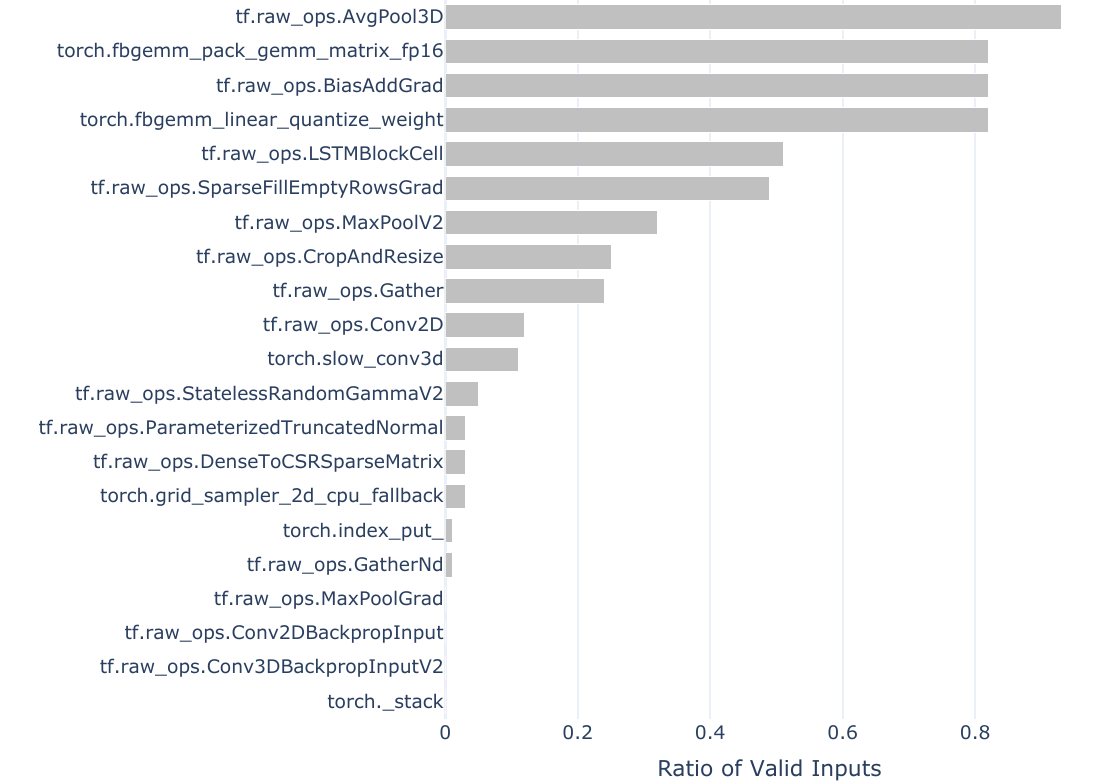}
    \caption{Ratio of valid inputs in the training data generated for
      the APIs with bugs.}
    \label{fig:acetest-buggy-ops-rq3}
  \end{figure}
  
  Finally, we assess how 
  the inclusion of the ML models affects 
  the bug finding capabilities of ACETest. 
  To do so, 
  we consider a set of bug-triggering inputs 
  previously reported by ACETest, and analyze 
  whether the ML models 
  correctly predict 
  the validity of these inputs.
  As in ACETest+ML 
  the models are used before 
  actually calling the target API,
  we expect the models to 
  correctly predict as valid 
  the bug-triggering inputs, 
  which would indicate that 
  ACETest+ML is also able to
  detect the bugs.
  
  For this analysis, we consider 
  40 bugs found by ACETest, 
  12 in \torch{} and 28 in \tf{}. 
  Note that, though ACETest's replication package~\cite{acetest2023} 
  includes 85 bugs, we only consider the bugs 
  related to the target operations
  and not in the API's validity checks.
  To ensure that our models are trained with values 
  in adequate ranges, we inspect all the bug-triggering inputs
  to collect ranges for int and float values,
  and maximum values for tensor shapes and dimensions,
  which we use when generating the training data.
  
  While our training data generation strategies are 
  useful for many APIs,
  they do not always guarantee the generation
  of positive samples, especially for APIs
  with complex constraints and various arguments.
  Considering the APIs related to the 40 bugs, 
  we were able to train
  ML models for 22 (55\%) of them. 
  For the remaining cases, 
  including complex APIs such as
  \CodeIn{torch.histogramdd}
  or \CodeIn{tf.raw\_ops.MaxPool3DGrad} 
  requiring several input tensors 
  with related shape values,
  we were not able to train the models. 
  Thus, we focus the experiment on the 22 APIs
  for which we were able to train the models, 
  and argue that with a relatively low 
  ratio of valid inputs in the training data (10\%),
  potentially obtained from 
  mining GitHub data, using an LLM, or even using FreeFuzz~\cite{Wei2022},
  we could train accurate models for 
  the remaining APIs.
  Figure~\ref{fig:acetest-buggy-ops-rq3} shows the 
  ratio of valid inputs in the training data 
  generated for the 22 APIs. 
  
  \begin{table}[t]
    \tabcolsep=0.14cm
    \centering
    \caption{Successful predictions on buggy inputs considering models 
    trained with valid-input ratios above a threshold.}
    \begin{tabular}{lrrrrrrrrr}
    \toprule
    Ratio >= & 0\% & 1\% & 5\% & 10\% & 20\% & 30\% & 40\% & 50\%  \\
    \midrule
    Success & 72\% & 77\% & 84\% & 91\% & 90\% & 100\% & 100\% & 100\% \\
    \bottomrule
    \end{tabular}
    \label{tab:ratio-valid-inputs-threshold-rq3}
  \end{table}
  
  Considering the 22 bugs, our models correctly 
  predict as valid 16 bug-triggering inputs (72\%).
  Note that for several APIs (e.g., 
  \CodeIn{tf.raw\_ops.MaxPoolGrad})
  the training data included a very low ratio 
  of valid inputs, insufficient to train
  accurate models. Thus, we investigate 
  the success rate of models trained with 
  different ratios of valid inputs.
  Table~\ref{tab:ratio-valid-inputs-threshold-rq3}
  shows the success rate when considering 
  models trained with a ratio of valid inputs
  greater than a threshold, 
  ranging from 0\% to 50\%.
  Notably, even from a small ratio 
  of 10\% of valid inputs,
  our models achieve a success rate $>$90\%, 
  being able to correctly predict as valid 
  the majority of the bug-triggering inputs.
  
  This analysis shows that including our models in the 
  ACETest pipeline has a negligible impact on
  its bug finding capabilities.

\section{Discussion}
\label{discussion}

\subsection{Threats to Validity}
An important threat to the validity of our study is the
randomness involved in the generation 
of the training data and 
in the training of the ML models, 
which may lead to different results
in each run.
To mitigate this threat, we repeat the training process
10 times for each target API,
each time using a different random seed. 
For each resulting dataset, the ML models are trained
and then the average performance is computed 
and reported in the results.

Another threat to the validity of our study is the 
assumption that the existing input validation code
is correct, which is fundamental for the training
process.
The rationale is
that these are important libraries used by lots of people; preventing
API misuse is very important.
So, it is a reasonable assumption to make
in this context. It is worth noting that
this prior work on testing \dl\ APIs also make that assumption~\cite{acetest2023,deepconstr2024}. 
To mitigate this threat, we implemented 
a new input validation code from the documentation 
of the API
for a random sample of 10 target APIs, 
and then compared the behavior with respect 
to the existing input validation code.
For these cases, 
we observed that the input validation code
is consistent with the documentation,
which gives us confidence on its correctness.

\subsection{Limitations}

\noindent\textbf{Data Generation for Training.}
Our study incorporates two strategies 
to automatically produce training data 
for each API under analysis: Random and Pairwise.
Although these two strategies are effective to generate
datasets from which we train highly-accurate ML models, conceptually, they may not be able to
produce a sufficient number of positive samples
for some operations, or even do not generate
positive samples at all,
preventing the training of the models. We remain to explore other strategies for training data generation
could be explored, such as 3-wise or 4-wise
combinations of the input arguments,
and even the use of standard
sampling techniques typically used in
the training of ML models.

\noindent\textbf{Data Generation for Testing.}  Our results
demonstrate that ML models can be used to improve the efficiency of
API-level fuzzers, such as ACETest~\cite{acetest2023}.  We encode the
problem of learning DL library constraints as a binary classification
problem, in which ML classifiers are trained to predict input validity. Our study shows that the integration of classifiers with
an existing test generator improves their performance. However, it is
important to note the fundamental limitation of this approach
considering the time wasted in generating inputs that the ML
classifiers will later discard. Note that generating inputs in batches
alleviates the problem.
An interesting direction of future work is to
explore the use of generational models (e.g.,
GANs~\cite{DBLP:journals/cacm/GoodfellowPMXWO20},
Transformers~\cite{vaswani2023attentionneed}, and Variational
Autoencoders~\cite{kingma2022autoencodingvariationalbayes}) to create
input data that is likely to be valid.


\section{Related Work}
\label{related-work}




\subsubsection*{Testing DL libraries}

The increasing advances in the development of
ML-based systems requires the availability of 
reliable DL libraries.
As a result, the testing of DL libraries has become
a very active research area~\cite{Phan_ETAL_ICSE2019, AUDEE_ASE20, LEMON_FSE20, 
muffin2022, nnsmith2023, acetest2023, Wei2022, 
deepconstr2024, docter2022, Deng_ETAL_ISSTA23, Deng_ETAL_FSE22, 
Liu_ETAL_ICSE23, Jia_ETAL_JSS21, LEMON_FSE20, deng2024large}.
These approaches can be divided into two categories: model-level fuzzers and api-level
fuzzers, with different ways of extracting constraints.

CRADLE~\cite{Phan_ETAL_ICSE2019} is a model-level fuzzer that takes 
pre-trained DL models as input and resolves the test oracle challenge 
with differential testing by comparing the inference results
from running the models on CPU vs GPU. Using existing models allows CRADLE 
to bypass input constraint checks.
AUDEE~\cite{AUDEE_ASE20} and LEMON~\cite{LEMON_FSE20} mutate
inputs and weights of existing models using different fitness functions
to derive new test input DL models. On the other hand, 
Muffin~\cite{muffin2022} generates new DL models by converting
the structure of a model to a computational flow graph and mutating
the substructures. NNSmith~\cite{nnsmith2023} takes API constraints as input
from the user and uses an SMT solver to generate valid DL models.

All of these techniques require API constraints to be provided as input. On the other hand,
NeuRI~\cite{Liu_ETAL_ICSE23} is a model-level fuzzer that can automatically derive API constraints
by instrumenting programs and invoking these programs
to inductively synthesize the operator rules. This process requires both valid and invalid
API invocations to infer constraints, which are not available for all DL library APIs. To combat this,
NeuRI mutates existing programs to generate the required diversity of valid and invalid API invocations.

FreeFuzz~\cite{Wei2022} is an API-level fuzzer that addresses type constraints on API parameters by collecting and executing open-source programs with DL library API calls to infer data types. Similarly, DocTer~\cite{docter2022} extracts these type constraints from documentation.
DeepREL~\cite{Deng_ETAL_FSE22} enhances FreeFuzz by identifying APIs with equivalent parameters and outputs, allowing it to generate test cases that involve multiple APIs using the same inputs. However, these techniques can still produce invalid inputs through mutation, as they do not account for tensor shape constraints of the API parameters.

Recently, Large Language Models (LLMs) based techniques have been introduced to test DL libraries.
TitanFuzz~\cite{Deng_ETAL_ISSTA23} leverages generative and infilling LLMs to generate input programs for testing DL libraries. FuzzGPT~\cite{deng2024large} 
provides existing bug reports to LLMs and asks it to generate partial or complete code snippets that can be used as test cases. 
Due to the inherent nature of LLMs, invalid inputs are still generated and these techniques also demand greater computational resources and time compared to traditional methods.

ACETest~\cite{acetest2023} leverages validity checks embedded in source code to automatically extract input constraints for DL APIs to generate valid test cases. DeepConstr~\cite{deepconstr2024} enhances existing constraints by identifying and expanding overly restrictive ones, improving the bug detection ability of the generated inputs.
The inputs generated by these techniques, however, exhibit low precision and recall, leading to a significant portion of invalid test cases for DL APIs.

In our study, we demonstrate how our trained models can enhance ACETest by incorporating a pre-filtering mechanism. This mechanism discards invalid inputs before executing the APIs. This approach can be replicated in other constraint-based methods in DL library testing (e.g. NeuRI) and other fuzzers, by integrating ML models to improve input validity prediction.

\subsubsection*{ML for Constraint Learning} 

The use of ML models to learn constraints 
in different domains has been studied in 
the literature~\cite{DBLP:conf/icse/BrunE04, Molina2019, Usman2019, Molina2022}.
Brun \etal~\cite{DBLP:conf/icse/BrunE04} studied the 
use of support vector machines and decision trees
to classify program properties that may lead to errors.
More recent approaches focus on 
learning constraints for complex data structure implementations 
in Java programs~\cite{Molina2019, Usman2019, Molina2022,Molina_ETAL_ICSE2022}.
For instance, Molina et al.~\cite{Molina2019}
proposed a technique based on artificial neural networks
to learn to distinguish between valid and invalid
input data structures for Java programs.
Similarly, Usman et al.~\cite{Usman2020}
studied the use of ML models 
to learn relational properties 
of data structures. 
In our study, we focus on learning constraints
for DL library operations, which requires the models 
to capture constraints related to tensor
dimensions and shapes. 

\section{Conclusion and Future Work} 
\label{conclusion}


Testing DL libraries is a very important 
task to ensure the reliability of DL applications.
Successfully testing DL library operations 
typically requires providing
inputs satisfying complex constraints 
imposed by these operations.
The difficulties in automatically generating 
such inputs have motivated the development
of techniques to infer the constraints
of DL library operations, and then use these constraints
to guide the generation of test cases.
However, these techniques still show limitations
in terms of false positives.


In this paper, we present a study on the ability of
state-of-the-art ML models to capture 
input constraints of operations 
from two popular DL libraries,
\torch{} and \tf{}.
Our intuition is that the availability of
input validation code in these libraries
can be exploited to generate training data
to produce ML classifiers that 
accurately predict the validity of inputs
for these operations.
We show that ML models can be 
very effective in classifying input validity
as measured by the traditional 
precision and recall metrics. 
Furthermore, we analyze the generalizability
of these models, and their potential to improve
the test input generation process for DL library operations.
Our results show that ML models generalize
well to unseen data with over 91\% accuracy,
and that they can be effectively
used to improve the pass rate of a 
state-of-the-art test input generation process,
increasing it from 29.1\% to 60.7\%.

As future work, we plan to explore two 
research directions. 
Firstly, we plan to investigate 
other training data generation strategies, 
such as 3-wise or 4-wise combinations of the input arguments,
that may allow us to support a wider range of operations. 
Secondly, we plan to address the 
test input generation problem with generative models,
such as GANs, to study the ability 
of such models to efficiently 
generate valid inputs for such operations.
We believe our study opens up
a promising research direction
to improve the testing of DL library operations
using state-of-the-art ML models.

\section*{Acknowledgments}
\sloppy
This work is partially supported by the U.S. National Science
Foundation under Grant Nos. CCF-2349961 and CCF-2319472, and
partially supported by the Ramón y Cajal fellowship RYC2020-030800-I,
by the Spanish Government through grants TED2021-132464B-I00
(PRODIGY), PID2022-142290OB-I00 (ESPADA), and CEX2024-001471-M funded
by MICIU/AEI/10.13039/501100011033 and by the Comunidad de Madrid as
part of the ASCEND project co-funded by FEDER Funds of the European
Union.


\balance

\clearpage
\bibliographystyle{ACM-Reference-Format}
\bibliography{main}

\end{document}

%% file: packages.tex
\setcitestyle{numbers,sort&compress}
\usepackage{amsmath}
\usepackage{color, soul, xcolor}  
\usepackage{colortbl}
\usepackage{booktabs} 
\usepackage{enumitem}
\usepackage{listings}
\usepackage{balance}
\usepackage{makecell}
\usepackage[normalem]{ulem} 
\usepackage{multirow}
\usepackage{tabularx}
\usepackage{longtable}
\usepackage{romannum}
\usepackage{textcomp}
\usepackage[T1]{fontenc}
\usepackage{newfloat}
\usepackage{tikz}
\usetikzlibrary{positioning}
\usetikzlibrary{tikzmark}
\usepackage[htt]{hyphenat}
\usepackage{float}
\definecolor{shadecolor}{gray}{0.97}
\usepackage[most]{tcolorbox}
\usepackage[linewidth=1pt]{mdframed}
\usepackage{array}
\usepackage{CJKutf8}
\usepackage{pifont}  
\usepackage{subcaption}
\usepackage{url}

\definecolor{keywordcolor}{rgb}{0.5, 0, 0.5}     
\definecolor{stringcolor}{rgb}{0.2, 0.6, 0.2}    
\definecolor{commentcolor}{rgb}{0.5, 0.5, 0.5}   
\definecolor{backgroundcolor}{rgb}{0.95, 0.95, 0.92} 

%% file: macros.tex
\newcolumntype{C}[1]{>{\centering\arraybackslash}p{#1}}
\usepackage{wrapfig}
\usepackage{pythonhighlight}
\newcommand{\CodeIn}[1]{{\small\textsc{#1}}}

\definecolor{mintbg}{rgb}{0.94, 1.0, 1.0} 

\newcommand{\MYROMAN}[1]{%
  \textup{\uppercase\expandafter{\romannumeral#1}}%
}
\newcommand{\Sum}[2]{\the\numexpr #1 + #2 \relax}
\newcommand{\Contrib}[1]{$\star$#1}
\newcommand{\etal}{et al.}
\newcommand{\ie}{i.e.}

\newcommand{\dl}{DL}

\newcommand{\nAPIpassrate}{41}
\newcommand{\ksPvalueACETestpassrate}{0} 
\newcommand{\ksPvalueACETestwMLpassrate}{0} 
\newcommand{\wilcoxonPvaluepassrate}{0.00408} 
\newcommand{\cohensDvaluepassrate}{0.76} 
\newcommand{\effectSizepassrate}{medium} 

\newcommand{\torch}{Pytorch}
\newcommand{\tf}{TensorFlow}

\newcommand{\freefuzz}{\textsc{FreeFuzz}}

\newcommand{\sota}{SoTA} 
